\documentclass[journal]{IEEEtran}

\usepackage{graphicx}
\graphicspath{ {./} }
\DeclareGraphicsExtensions{.pdf,.jpeg,.jpg,.png}

\ifCLASSOPTIONcompsoc
    \usepackage[caption=false, font=normalsize, labelfont=sf, textfont=sf]{subfig}
\else
\usepackage[caption=false, font=footnotesize]{subfig}
\fi

\usepackage{amsmath}
\usepackage{physics}
\usepackage{xcolor}
\usepackage[super]{nth}

\usepackage{cases}
\usepackage{cite}

\newcommand{\etal}{\textit{et al}.}

\usepackage{algorithm}
\usepackage{algorithmic}

\begin{document}
%
\title{A Control Chart Approach to Power System Line Outage Detection Under Transient Dynamics}

\author{Xiaozhou~Yang,~\IEEEmembership{Student Member,~IEEE}%
,~Nan~Chen,~\IEEEmembership{Member,~IEEE}
,~and~Chao~Zhai,~\IEEEmembership{Member,~IEEE}
\thanks{This work is partially supported by the Future Resilient Systems programme at the Singapore-ETH Centre, which was established collaboratively between ETH Zurich and Singapore's National Research Foundation (FI 370074011) under its Campus for Research Excellence and Technological Enterprise programme.}%
\thanks{X. Yang and Nan Chen are with the Department
of Industrial Systems Engineering and Management, National University of Singapore, 117576 Singapore (e-mail: xiaozhou.yang@u.nus.edu; isecn@nus.edu.sg). X. Yang is also with ETH Zurich, Future Resilient Systems, Singapore-ETH Centre, 1 CREATE Way, CREATE Tower, 138602 Singapore}
\thanks{Chao Zhai is with the School of Automation, China University of Geosciences, Wuhan 430074 China, and with Hubei Key Laboratory of Advanced Control and Intelligent Automation for Complex Systems, Wuhan 430074, China (email: zhaichao@amss.ac.cn).}
}


%


\maketitle

\begin{abstract}
Online transmission line outage detection over the entire network enables timely corrective action to be taken, which prevents a local event from cascading into a large scale blackout. Line outage detection aims to detect an outage as soon as possible after it happened. Traditional methods either do not consider the transient dynamics following an outage or require a full Phasor Measurement Unit (PMU) deployment. Using voltage phase angle data collected from a limited number of PMUs, we propose a real-time dynamic outage detection scheme based on alternating current (AC) power flow model and statistical change detection theory. The proposed method can capture system dynamics since it retains the time-variant and nonlinear nature of the power system. The method is computationally efficient and scales to large and realistic networks. Extensive simulation studies on IEEE 39-bus and 2383-bus systems demonstrated the effectiveness of the proposed method.
\end{abstract}


\section{Introduction}
\IEEEPARstart{W}{ith} the emergence and integration of distributed energy resources, there is increasing volatility in modern power systems. Ensuring a reliable electricity supply is an essential but challenging task for independent system operators (ISOs). To this end, fast anomalous event detection is necessary to contain system disruptions and minimize the potential impact. Power systems can experience numerous types of disruptions. Among them, power line outage receives a significant amount of attention from both the research community and industry. Power line outages can happen due to reasons like adverse weather conditions or component degradation. An outage, if not detected and addressed in time, could lead to severe disruptions and possibly cascading failures.

Real-time situational awareness about the system, e.g., changes in operating conditions and external system contingencies, enables ISOs to promptly identify and respond to abnormal events \cite{kezunovic2013role}. Without it, a local disruption can cascade into a large-scale blackout. One of the common contributing factors of the 2003 Northeast and 2011 Southwest blackout was that the ISOs were not alerted in time about external outage contingencies, e.g., tripping of a critical transmission line \cite{Johnson2000}. One of the challenges about real-time monitoring is that outage dynamics can manifest in a time scale of milliseconds \cite{Milano2010power}. Traditional supervisory control and data acquisition (SCADA) system is not able to capture these dynamics since it reports at a rate of one measurement every several seconds \cite{pignati2015real}.
On the other hand, the increasing penetration of Phasor Measurement Units (PMUs) makes many real-time monitoring, protection, and control applications possible. PMUs are devices installed at substations capable of recording high-fidelity GPS time-synchronized phasors. An industry-grade PMU can measure voltage and current phasors on the bus with a total vector error of less than $1\%$, and with a reporting rate of 30 to 60 samples per second. As an essential part of the wide-area management system, many consider PMU technology as the key to grid modernization. PMU technologies are actively studied for tasks such as power oscillation monitoring \cite{Abdi-Khorsand2017}, abnormal event detection \cite{Mohamed2018,Kim2018}, and dynamic state and parameter estimation \cite{Chakrabortty2009,Chavan2017}. For a comprehensive review of PMU applications in the power system, readers can refer to \cite{Aminifar2014}.


There is a growing body of work on power line outage detection leveraging on PMU data. Many of the detection schemes focus on the monitoring of bus voltage phase angles. They are implemented in real-time, taking advantage of the high-reporting rate of PMUs. Consequently, a common challenge is to keep the computational cost low while still making sure that useful information can be extracted. Another challenge is that not all buses are equipped with a PMU device, making some parts of the system unobservable. Current related works of line outage detection using PMU data can be classified by the two approaches taken. One is a data-driven approach where no or very little physical knowledge about the system is required \cite{Xie2014, Rafferty2016, Hosur2019}. On the other hand, many take a hybrid approach where first-principle models are incorporated with data-driven methods \cite{Jamei2016, Jamei2017a, ardakanian2017event, Ardakanian2019a, Tate2008, Chen2016, Rovatsos2017}. 

\subsubsection{Data-driven Approach}
Using principal component analysis (PCA), Xie $\etal$ monitor the reconstruction error of PMU measurements using a lower-dimensional representation obtained from data under an outage-free condition\cite{Xie2014}. 
Similarly, using PCA of frequency measurements, Rafferty $\etal$ design a control chart to detect and classify abnormal frequency events\cite{Rafferty2016}. 
Hosur and Duan construct an observation matrix under a normal condition by modeling the network as a linear time-invariant system \cite{Hosur2019}. An alarm is raised whenever the underlying null space of the observation matrix changes. The method requires a window of samples to reflect a null space change.
Without a physical model, these data-driven schemes are flexible enough to detect both outages and other abnormal events. However, they often face difficulties when the events have a low signal-to-noise ratio, e.g., outages with mild phase angle disturbances. The hybrid approach, on the other hand, augments PMU data with physical system information to improve the detection performance under such conditions. 

\subsubsection{Hybrid Approach}
Using Ohm's law, Jamei $\etal$ show that the correlation matrix between voltage and current measurements of a pair of buses has rank one under normal condition \cite{Jamei2017a}. An alarm is raised once this dependency changes. However, currents and voltages at both ends of the line are required. 
Another group of work assumes that the power system settles into a quasi-steady state immediately after an event. Ardakanian $\etal$ monitor the discrepancy between measured and computed steady-state currents using recovered admittance matrix \cite{Ardakanian2019a}. Using pre- and post-outage steady-state bus angles, outage detection is formulated as an optimization problem by Tate and Overbye \cite{Tate2008} and a quickest change detection problem by Chen $\etal$ \cite{Chen2016}. This line of work does not require all buses to be monitored by a PMU. However, the steady-state approximation would not be sufficient at describing the actual system behavior following an outage. A later work by Rovatsos $\etal$ attempts to incorporate transient dynamics using pre-determined participation factors (PF) matrix \cite{Rovatsos2017}. However, the adequacy of the chosen PF matrix at describing system response following an outage is not studied. 
 
To the best of the authors' knowledge, there is minimal work on detection schemes that allow unobservable buses and consider system transient response to an outage. In this work, we take a hybrid approach where the power system model is the basis for the statistical detection method. We derive a time-variant small-signal relationship between net active power and nodal voltage phase angles from the AC power flow model. Outage detection is then formulated as a statistical distribution change detection problem. A generalized likelihood ratio (GLR) detection scheme is implemented to detect the outage at a pre-specified false alarm rate. 

The main contributions of our work can be summarized in two aspects. Firstly, our power system model retains the non-linear and time-varying characteristics of system transient response that follows after the outage. The system is not assumed a quasi-steady state immediately after the disruption. In fact, from our dynamic outage simulation, we observed that the transient response could last over 10 seconds. Secondly, the proposed GLR detection scheme can deal with the trade-off between system-wide false alarm rate and detection delay. The ability to decide among different detection thresholds gives ISOs the flexibility to cater to their system needs. The detection scheme is also computationally efficient, therefore friendly for online implementation in a large network.

The remainder of this paper is organized as follows. Section \ref{sec:formulation} describes the power system model and the statistical model used to characterize system behaviors before and after the outage. Then, dynamic detection and identification scheme is developed in Section \ref{sec:detection_scheme}. Effectiveness of the proposed scheme on simulation data of two test power systems are reported and discussed in Section \ref{sec:results}. In Section \ref{sec:conclusion} we conclude the paper with two future research directions.

\section{Problem Formulation}
\label{sec:formulation}
\subsection{Power System Model}
\label{sec:power_model}
Given a power network where $N$ buses are connected by $L$ power lines, power flows in the network can be characterized by a set of non-linear algebraic equations called the AC power flow equations. This set of equations describes the relationship between net active power (P), net reactive power (Q), nodal voltage magnitude (V), and voltage phase angle ($\theta$) governed by Kirchhoff's circuit laws. They can be written as:
\begin{subequations}
\label{eqn:AC_power_flow_model}
\begin{align}
\text{P}_m &= \text{V}_m \sum_{n=1}^{N} \text{V}_n \text{Y}_{mn} \cos (\theta_m - \theta_n - \alpha_{mn}) \,, \label{eqn:AC_power_flow_P}\\
\text{Q}_m &= \text{V}_m \sum_{n=1}^{N} \text{V}_n \text{Y}_{mn} \sin (\theta_m - \theta_n - \alpha_{mn}) \,, \label{eqn:AC_power_flow_Q}
\end{align}
\end{subequations}
for bus $m = 1, 2, \dots, N$ \cite{Glover2012}. Y$_{mn}$ is the magnitude of the $(m,n)_{th}$ element of the bus admittance matrix $\boldsymbol{Y}$ when the complex admittance is written in the exponential form, i.e. 
\begin{equation}
\label{eqn:admittance_expression}
\text{Y}_{mn}e^{j\alpha_{mn}} = G_{mn} + jB_{mn} \,.
\end{equation}
For a bus equipped with PMU, V and $\theta$ are measured and available. We also assume that elements of the bus admittance matrix are known. For a large system, $\boldsymbol{Y}$ is usually a sparse matrix since any single bus only has a few incident buses, i.e., Y$_{mn} = 0$ if bus $m$ and bus $n$ are not connected. The system topology is embedded in the admittance matrix $\boldsymbol{Y}$. In particular, the admittance matrix is constructed by 
\begin{equation}
\label{eqn:admittance_matrix}
\boldsymbol{Y} = \mathbf{A [y] A}^{T}
\end{equation} where $\mathbf{A}$ is the bus to branch incidence matrix with columns representing lines and rows as buses. $\mathbf{A}^T$ is the transpose of $\mathbf{A}$. For the $l_{th}$ line transmitting power from bus $m$ to bus $n$, the $l_{th}$ column of the matrix $\mathbf{A}$ has 1 and -1 on the $m_{th}$ and $n_{th}$ position and 0 everywhere else. $\mathbf{[y]}$ is the diagonal matrix with individual line admittances on the diagonal.

Without the loss of generality, we assume bus 1 is the reference bus. This bus serves as the angular reference to all other buses, and its phase angle is set to $0^\circ$. The voltage magnitude at the reference bus is also set to $1.0$ per unit (p.u.). Let \textbf{P}, \textbf{Q}, $\boldsymbol{\theta}$, and \textbf{V} represent the $(N-1)$-dimensional column vectors of net active power, net reactive power, voltage angles and magnitudes respectively at all buses except the reference bus. Taking a derivative with respect to time $t$ on both sides of (\ref{eqn:AC_power_flow_model}), we obtain
\begin{equation}
\label{eqn:ac_jacobian}
\left[\begin{array}{c} \frac{\partial \textbf{P}}{\partial t} \\[.5em] \frac{\partial \textbf{Q}}{\partial t}\end{array}\right]
=
\left[\begin{array}{c|c}
\mathbf{J}_1 & \mathbf{J}_2 \\
 \hline \mathbf{J}_3 & \mathbf{J}_4 
 \end{array}\right] 
\cdot \left[\begin{array}{c} \frac{\partial \boldsymbol{\theta}}{\partial t} \\[.5em] \frac{\partial \textbf{V}}{\partial t}\end{array}\right] \,,
\end{equation}
where $\mathbf{J}_i, i = 1, \dots, 4$ are the four submatrices of the AC power flow Jacobian with 
\begin{equation}
\mathbf{J}_1 = \frac{\partial \textbf{P}}{\partial \boldsymbol{\theta}} \,,  \mathbf{J}_2 = \frac{\partial \textbf{P}}{\partial \mathbf{V}} \,, \mathbf{J}_3 = \frac{\partial \textbf{Q}}{\partial \boldsymbol{\theta}} \,, \mathbf{J}_4 = \frac{\partial \textbf{Q}}{\partial \textbf{V}} \,.
\end{equation}
In the usual operating range of relatively small angles, real power systems exhibit much stronger interdependences between \textbf{P} and $\boldsymbol{\theta}$ and between \textbf{Q} and \textbf{V} than those between \textbf{P} and \textbf{V} and between \textbf{Q} and $\boldsymbol{\theta}$ \cite{murty2017power}. By neglecting $\mathbf{J}_2$ and $\mathbf{J}_3$, (\ref{eqn:ac_jacobian}) reduces to the decoupled AC power flow equations where the changes in voltage angles and magnitudes are not coupled, i.e. 
$
\mathbf{J}_2 = \mathbf{J}_3 = \mathbf{0}.
$
Therefore, we obtain a small-signal time-variant model describing the relationship between active power mismatches and the changes in voltage angles: 
\begin{equation}
\label{eqn:ac_decoupled_jacobian}
\frac{\partial \textbf{P}}{\partial t} \approx \mathbf{J}_1(\boldsymbol{\theta}) \frac{\partial \boldsymbol{\theta}}{\partial t}\,.
\end{equation}
From here onwards, we drop the subscript $1$ from $\mathbf{J}_1$. The off-diagonal and diagonal elements of the $\mathbf{J}$ matrix can be derived from Eqn (\ref{eqn:AC_power_flow_P}) respectively:
\begin{subequations}
\label{eqn:elements_J}
\begin{align}
	\frac{\partial \text{P}_{m}}{\partial \theta_{n}} 
	& = \text{V}_{m} \text{V}_{n} \text{Y}_{m n} \sin \left( \theta_{m} - \theta_{n} - \alpha_{m n} \right) \,,  m \neq n \,,\label{eqn:elements_J_off}\\ 
	\frac{\partial \text{P}_{m}}{\partial \theta_{m}} 
	& = -\sum_{ n=1 \atop n \neq m}^{N} \text{V}_{m} \text{V}_{n} \text{Y}_{m n} \sin \left( \theta_{m} - \theta_{n} - \alpha_{m n} \right) \,. \label{eqn:elements_J_diag} 
\end{align}
\end{subequations}
Note that $t \in [0, \infty)$ is implicit in the continuous-time quantities $\textbf{P}, \textbf{V}$ and $\boldsymbol{\theta}$. Accordingly, we define their discrete counterparts as $\textbf{P}_k, \textbf{V}_k$ and $\boldsymbol{\theta}_k$ at time $t_k$ for $k = 1, 2, \dots$. For PMU devices with a sampling frequency of 30 Hz, $\Delta t = t_{k} - t_{k-1} = 1/30$ s. A first-order difference discretization by Euler's formula can approximate (\ref{eqn:ac_decoupled_jacobian}) by:
\begin{equation}
\label{eqn:small_signal_model}
    \Delta \textbf{P}_k =  \mathbf{J}(\boldsymbol{\theta}_{k-1}) \Delta \boldsymbol{\theta}_k \,,
\end{equation}
where $\Delta \textbf{P}_k = \textbf{P}_k - \textbf{P}_{k-1}$ and $\Delta \boldsymbol{\theta}_k = \boldsymbol{\theta}_k - \boldsymbol{\theta}_{k-1}$, i.e. the active power mismatch and difference between two consecutive angle measurements. We have derived a time-variant relationship between variations in phasor angles and net active power on buses. 
The key feature of our model lies in the $\mathbf{J}$ matrix in (\ref{eqn:small_signal_model}). The matrix changes with $\boldsymbol{\theta}$, which in turn changes with time. Therefore, it retains the non-linear and dynamic nature of the AC power system.  

\begin{figure}[!t]
\centering
\includegraphics[width=1\linewidth]{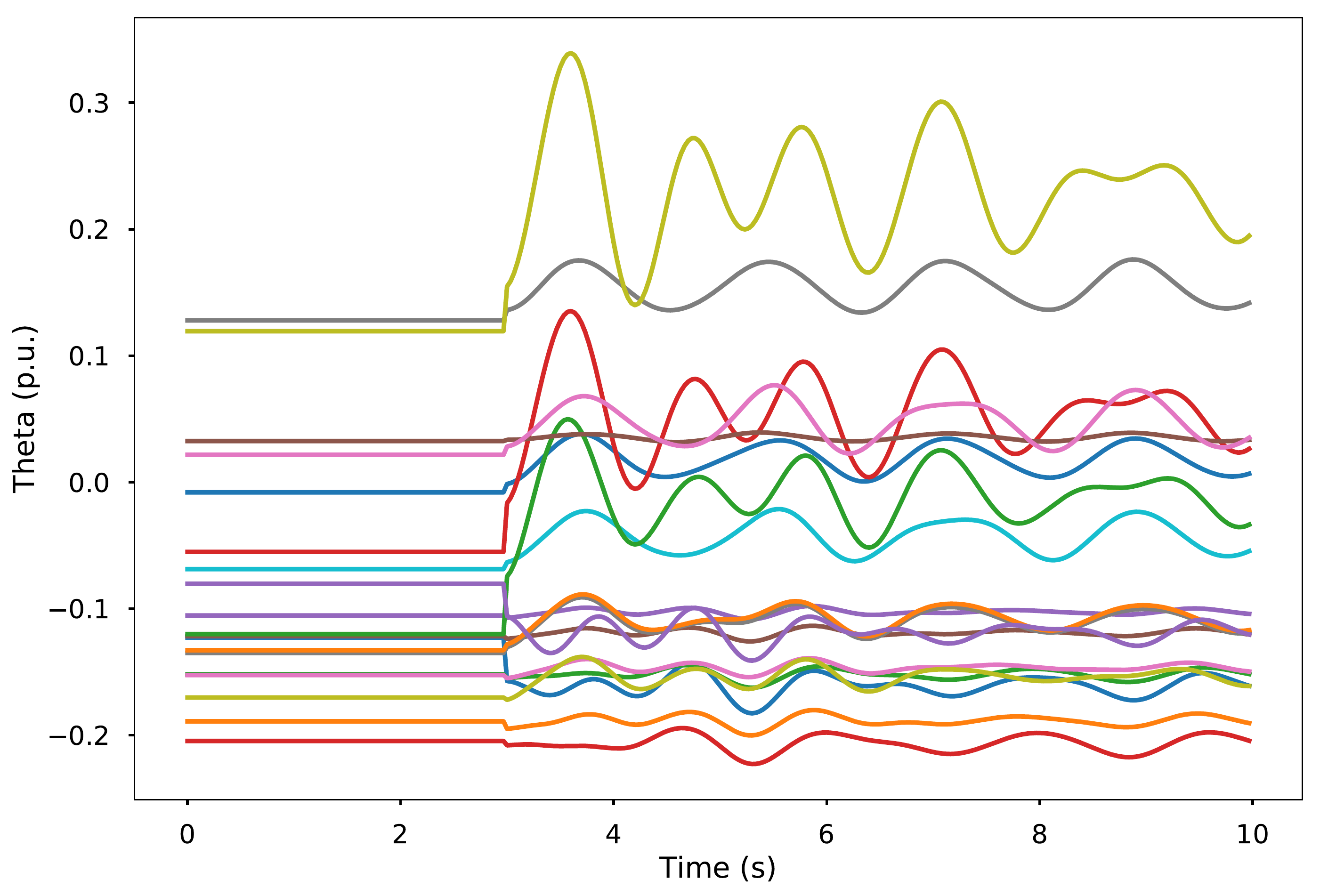}
\caption{\textit{The progression of bus voltage phase angles after an outage at $t=3$ s, where each line corresponds to one bus. The steady-state bus angle balance is severely distorted during the transient response phase.}}
\label{fig:severe_transients}
\end{figure}

Methods relying on a static relationship between $\Delta\textbf{P}$ and $\Delta\boldsymbol{\theta}$ make three further assumptions about the system \cite{Tate2008, Chen2016}: 1) flat voltage profile, i.e. V$_m \approx$ V$_n \approx 1.0$ p.u.; 2) approximately homogeneous bus angles across the network, i.e. $\cos(\theta_m - \theta_n) \approx 1, \sin(\theta_m - \theta_n) \approx 0$; 3) reactive property of a line is much more significant than its resistive property, i.e. $B_{mn} \gg G_{mn}$. Under these assumptions, (\ref{eqn:ac_decoupled_jacobian}) reduces to
\begin{equation}
\label{eqn:dc_power_flow}
\frac{\partial \textbf{P}}{\partial t} \approx -\mathbf{B} \frac{\partial \boldsymbol{\theta}}{\partial t} \,,
\end{equation}
where $\mathbf{B}$ is the imaginary component of $\boldsymbol{Y}$. 
While line resistances in transmission systems are generally one order of magnitude smaller than reactances, this is not usually the case for distribution systems \cite{anderson1983stability}. Also, a static model may not be accurate enough to reflect the transient behavior after an outage since the homogeneous angles assumption might be violated \cite{kaye1984analysis}. We routinely encounter this phenomenon in our dynamic simulation. For example, in Fig. \ref{fig:severe_transients}, the balance between voltage angles is severely distorted following an outage, e.g., at around $t = 3.75$ s.
Furthermore, the duration of transient dynamics is non-negligible for real-time detection purposes. Therefore, to reflect the dynamic behavior in a timely and accurate manner, $\mathbf{J}$ matrix in (\ref{eqn:small_signal_model}) is updated by real-time streaming PMU data.

\subsection{Statistical Model}
\label{sec:statistical_model}
For a balanced steady-state power system with no active power mismatch, we have $\textbf{P}_{0} = 0$. Within a short period of time, net active power fluctuates around zero as the generators respond to random changes in electricity demand. Therefore, we can model the trajectory of $\textbf{P}$ as a Brownian motion with drift $\boldsymbol{0}$ and variance $\sigma^2 t \mathbf{I}$ which is a continuous-time stochastic process: $\{\textbf{P}_t: t \in [0, \infty)\}$. $\sigma^2$ is pre-determined and $\mathbf{I}$ is an identity matrix of appropriate dimension. One of the implications of a Brownian motion is that their independent increment, i.e. $\Delta\textbf{P}_k = \textbf{P}_{t_k} - \textbf{P}_{t_{k-s}}$, follows a multivariate Gaussian distribution with mean $\boldsymbol{0}$ and variance $\sigma^2(t_k - t_{k-s})\mathbf{I}$ \cite{cox2017theory}. In particular, taking $s = 1$, we have 
$
t_{k} - t_{k-s} = \Delta t
$ and 
\begin{equation}
\Delta\textbf{P}_k \sim \mathcal{N}(\boldsymbol{0}, \sigma^2 \Delta t \mathbf{I}) \,.
\end{equation} Since $\sigma^2$ is pre-determined, we can replace $\sigma^2 \Delta t$ by $\sigma^2$ for notational simplicity. Rearranging the variables in (\ref{eqn:small_signal_model}), we have 
\begin{equation}
\Delta \boldsymbol{\theta}_k = \mathbf{J}(\boldsymbol{\theta}_{k-1})^{-1} \Delta \textbf{P}_k \,.
\end{equation}
Therefore, we can characterize bus angle variations by 
\begin{equation}
\label{eqn:angle_distribution}
    \Delta\boldsymbol{\theta}_k \sim  \mathcal{N}(\boldsymbol{0}, \sigma^2 (\mathbf{J}(\boldsymbol{\theta}_{k-1})^{T} \mathbf{J}(\boldsymbol{\theta}_{k-1}))^{-1}) \,.
\end{equation}

From (\ref{eqn:angle_distribution}), we see that the angle variations at time $k$ are characterized by the structure of $\mathbf{J}$ and the angle values at $t = k-1$. Let $\mathcal{L}$ represent the set of all possible combinations of outages, e.g., single-line outage, double-line outage. When an outage $\ell \in \mathcal{L}$ happens, the grid topology and the bus admittance matrix changes. The new bus admittance matrix $\boldsymbol{Y}_{\ell}$ induces a new $\mathbf{J}_{\ell}$, and therefore, a new distribution of $\Delta\boldsymbol{\theta}_k$. There is a one-to-one correspondence between an outage scenario and a distribution of $\Delta\boldsymbol{\theta}_k$. Furthermore, we assume that the outage is persistent, i.e., tripped lines are not restored in the time under consideration. We also assume that the outage would not result in any islanding in the network, i.e., no part of the system is isolated from the main grid. 

In light of the above characterization, we adopt a hypothesis testing framework to detect the distribution change in $\Delta\boldsymbol{\theta}_k$:
\begin{subequations}
\label{eqn:pre_post_distribution}
\begin{align}
 	 H_0: \Delta\boldsymbol{\theta}[k] &\sim \mathcal{N}(\boldsymbol{0}, \sigma^2 (\mathbf{J}_0^T \mathbf{J}_0)^{-1}) \,,  \label{eqn:pre_distribution}\\
 	 H_1: \Delta\boldsymbol{\theta}[k] &\sim \mathcal{N}(\boldsymbol{0}, \sigma^2 (\mathbf{J}_{\ell}^T \mathbf{J}_{\ell})^{-1}) \,, \ell \in \mathcal{L} \,, \label{eqn:post_distribution}
\end{align}
 \end{subequations}
for $k = 1, 2, \dots$. The null hypothesis is that there is no outage, and the corresponding Jacobian is $\mathbf{J}_0$. The alternative hypothesis is that there is an outage scenario $\ell$, where the corresponding Jacobian is $\mathbf{J}_{\ell}$. If we reject the null hypothesis at time $\tau$, then the distribution of $\Delta\boldsymbol{\theta}[k]$ has changed, and the outage is detected. The detailed procedure of real-time detection under this framework is described in Section \ref{sec:detection_scheme}.


A common challenge for PMU applications is that not all buses are equipped with a PMU. Here we adapt the previous formulations to a limited PMU deployment. Suppose $K$ PMUs are installed where $K < N$. Given a selection matrix $\mathbf{S} \in \{ 0 ,1 \}^{(K \times N)}$ that selects $K$ observable buses from the complete set of $N$ buses, observable bus angle data is 
\begin{equation}
\boldsymbol{\theta}^o_k = \mathbf{S} \boldsymbol{\theta}_k \,,
\end{equation} where $\mathbf{S}$ is a diagonal matrix of size $(K \times N)$ and entries equal to 0 or 1. The corresponding angle variations and Jacobian matrix are 
\begin{eqnarray}
\Delta\boldsymbol{\theta}^o_{k} &=& \mathbf{S} \Delta\boldsymbol{\theta}_k \,, \\
\mathbf{J}^o(\boldsymbol{\theta}^o_{k-1}) &=& \mathbf{S} \mathbf{J}(\boldsymbol{\theta}^o_{k-1}) \mathbf{S}^T \,.
\end{eqnarray}
Therefore, $\Delta\boldsymbol{\theta}^o_{k}$ is a $K$-dimensional vector and $\mathbf{J}^o(\boldsymbol{\theta}^o_{k-1})$ is a $(K \times K)$-dimensional matrix. To obtain the hypothesis testing framework in (\ref{eqn:pre_post_distribution}), we replace $\Delta\boldsymbol{\theta}_k, \mathbf{J}_0, \text{and } \mathbf{J}_{\ell}$ by $\Delta\boldsymbol{\theta}^o_k, \mathbf{J}^o_0, \text{and } \mathbf{J}^o_{\ell}$ respectively. 

\renewcommand{\IEEEQED}{\IEEEQEDoff}
\begin{IEEEproof}[Remark 1 (Setting up Outage Scenarios)]
The one-to-one correspondence between the Jacobian and grid topology can be established by looking at how the admittance matrix is constructed in (\ref{eqn:admittance_matrix}). $\boldsymbol{Y}$ is constructed from the bus incidence matrix $\mathbf{A}$ and the line admittances. For different outage scenarios, we just need to set the corresponding column of $\mathbf A$ to 0. For example, to set up the $l_{th}$ line outage, we set the entries in the $l_{th}$ column of $\mathbf{A}$ to 0 to get $\mathbf{A}_\ell$. The corresponding bus admittance matrix $\boldsymbol{Y}_{\ell}$ is obtained by $\boldsymbol{Y}_{\ell} = \mathbf{A}_\ell \mathbf{[y]} \mathbf{A}_\ell^{T}$. The Jacobian matrix $\mathbf{J}_{\ell}$ describing the post-outage system is obtained by (\ref{eqn:elements_J}). Therefore, no simulation or real data is needed to generate the outage scenarios to set up the monitoring scheme during offline preparation. In real applications, both the bus incidence matrix and the line admittances can be obtained based on the network topology and data during the outage-free period. It will then be sufficient to apply the proposed method.
\end{IEEEproof}

\begin{IEEEproof}[Remark 2 (Inaccuracy of Jacobian Due to Unobservable Neighbor Buses)]
For a limited PMU deployment, there may be some inaccuracies in the computed diagonal elements of $\mathbf{J}^o(\boldsymbol{\theta}^o_{k-1})$. In particular, if there is no PMU on bus $n$, a neighbor of bus $m$, measurements $\text{V}_{n}$ and $\theta_{n}$ would not be available. Therefore, the term, $-\text{V}_{m} \text{V}_{n} \text{Y}_{m n} \sin \left(\theta_{m}-\theta_{n}-\alpha_{m n}\right)$, would not be computable and is treated as 0 for the summation in (\ref{eqn:elements_J_diag}). The issue could be alleviated by carefully designing the PMU placement (locations). One possible design rule is to make sure that each observable bus has at least one observable neighbor bus. In general, PMU locations will influence the efficiency of outage detection. It is also of interest to practitioners to find the optimal placement of PMUs so that even with limited PMUs, we can detect outages as quickly as possible. However, the placement problem is beyond the scope of this paper, and we will study this topic in our future research. 
\end{IEEEproof}

\section{Outage Detection Scheme}
\label{sec:detection_scheme}
We have formulated the outage detection as a problem of distribution change detection under a hypothesis testing framework in Section \ref{sec:statistical_model}. In general, under normal conditions, system outputs follow a common distribution with a probability density function $f_0$. At some unknown time $\tau$, the system condition changes, and the density function changes to $f_1$. We wish to design a scheme where an alarm is raised once a monitoring statistic $W(\cdot)$ crosses a pre-defined threshold of $c$. The two key design aspects are: 1) how to compute the monitoring statistic, $W(\cdot)$; 2) how to determine the detection threshold, $c$. The monitoring statistic will be close to zero under a normal condition and increase unboundedly if a change happens. The detection threshold needs to be specified to meet a particular false alarm rate constraint. 

We adopt a GLR approach originally proposed by \cite{lorden1971procedures} to design the detection scheme. The scheme repeatedly evaluates the likelihood of a normal condition against the likelihood of an abnormal condition. In our problem, bus angle variations are not independent samples since the distribution at time $k$ is influenced by bus angles at time $k-1$ as shown in (\ref{eqn:angle_distribution}). However, $\Delta\boldsymbol{\theta}_k$ can be regarded as a conditionally independent random variable with density function $f_0(\cdot | \boldsymbol{\theta}_{k-1})$ under $H_0$ in (\ref{eqn:pre_distribution}) and, after an outage, with density function $f_{\ell}(\cdot | \boldsymbol{\theta}_{k-1})$ under $H_1$ in (\ref{eqn:post_distribution}). For every new data $\Delta\boldsymbol{\theta}_k$, we test $H_0$ against $H_1$ for some outage scenario $\ell \in \mathcal{L}$ using a log-likelihood ratio test statistic. In particular, let 
\begin{equation}
\label{eqn:log_likelihood_ratio}
Z_k(\ell) = \ln \frac{f_{\ell}(\Delta\boldsymbol{\theta}_{k} | \boldsymbol{\theta}_{k-1})}{f_{0}(\Delta\boldsymbol{\theta}_{k} | \boldsymbol{\theta}_{k-1})} \,
\end{equation} be the log-likelihood ratio of an outage scenario $\ell$ at time k. $Z_k(\ell)$ is positive if the likelihood of a change is larger than that of a normal condition. Then the test statistic is:
\begin{equation}
\label{eqn:glr_statistic_direct}
G_k = \max \left\lbrace  0, \, \underset{1\le i \le k}{\max} \, \underset{\ell \in \mathcal{L}}{\max} \sum_{j=i}^{k} Z_{j}(\ell)  \right\rbrace\,.
\end{equation}
and the GLR detection scheme will raise an alarm at the time:
\begin{equation}
\label{eqn:glr_detection_rule}
D = \inf \left\lbrace  k \ge 1: G_k \ge c \right\rbrace \,.
\end{equation}
Since the time and location of the outage are not known a priori, they are replaced by their maximum likelihood estimates. Schemes of the form involving searching through the maximum over time ($1\le i \le k$) and over likelihood ($\sum_{j=i}^{k} Z_{j}(\ell)$) are referred to as the GLR schemes. Such schemes have optimal properties in terms of their detection performance. Let $E_{H_0}(D)$ be the expectation of time of alarm when there is no outage, i.e., mean time to a false alarm. Suppose c is chosen such that the scheme satisfies a certain false alarm rate, $E_{H_0}(D) \ge \gamma\{ 1 + o(1) \}$. For conditionally independent data, Lai has proved that the detection rule (\ref{eqn:glr_detection_rule}) is asymptotically optimal in the sense that among all rules \(T\) with \(E_{H_0}(T) \geq \gamma\{ 1 + o(1)\}\), it minimizes the worst-case detection delay as defined by
\begin{equation}
\overline{E}_{H_1}(T)=\sup _{\tau \geq 1} \operatorname{ess} \sup E^{(\tau)}\left[(T-\tau+1)^{+} | \boldsymbol{\theta}_{1}, \cdots, \boldsymbol{\theta}_{\tau-1}\right] \,,
\end{equation} as the outage time $\tau \to \infty$ \cite{Lai1998}.

For the actual online implementation, we use an recursive formulation of the GLR scheme. Note that $G_k$ in (\ref{eqn:glr_statistic_direct}) can be rewritten as 
\begin{eqnarray}
G_k	&=& \max \, \left\lbrace  0,  \, \underset{\ell \in \mathcal{L}}{\max} \underset{1\le i \le k}{\max} \sum_{j=i}^{k} Z_{j}(\ell)  \right\rbrace\,, \nonumber\\
	&=& \underset{\ell \in \mathcal{L}}{\max} \, \max \left\lbrace  0,  \, \underset{1\le i \le k}{\max} \sum_{j=i}^{k} Z_{j}(\ell)  \right\rbrace\,,\nonumber \\
	&=& \underset{\ell \in \mathcal{L}}{\max}  \,  W_{\ell, k} \,.
\end{eqnarray} 
where in the first step we have switched the position of the two inner $\max$ operators since the overall maximum is not affected \cite{Mei2010}. Also, in the last step,
\begin{equation}
\label{eqn:glr_statistics}
W_{\ell, k} = \max \left\lbrace  0, W_{\ell, k-1} + Z_k(\ell) \right\rbrace \,,
\end{equation}
an equivalent recursive form of the term $\underset{1\le i \le k}{\max} \sum_{j=i}^{k} Z_{j}(\ell) $ in $G_k$. Therefore, for every scenario $\ell$, we just need to keep track of the monitoring statistic $W_{k-1}$ at the previous time step and obtain the log-likelihood ratio $Z_k$ at the current time step. $Z_k(\ell)$ can be found analytically by
\begin{equation}
\label{eqn:log_likelihood_explicit}
Z_k(\ell) = \ln\left| \mathbf{J}_\ell \right| - \ln\left| \mathbf{J}_0 \right| + \frac{1}{2\sigma^2} \Delta\boldsymbol{\theta}_k^{T} \left[ \mathbf{J}_0{^T}  \mathbf{J}_0 - \mathbf{J}_\ell{^T} \mathbf{J}_\ell \right] \Delta\boldsymbol{\theta}_k \,,
\end{equation} based on the multivariate Gaussian distribution likelihood function. Using the recursive formulation, the stopping time is
\begin{equation}
\label{eqn:stopping_rule}
    D = \inf \left\lbrace  k \ge 1: \underset{\ell \in \mathcal{L}}{\max} \, W_{\ell, k} \ge c \right\rbrace \,.
\end{equation}

\begin{figure}[!t]
\centering
\includegraphics[width=1\linewidth]{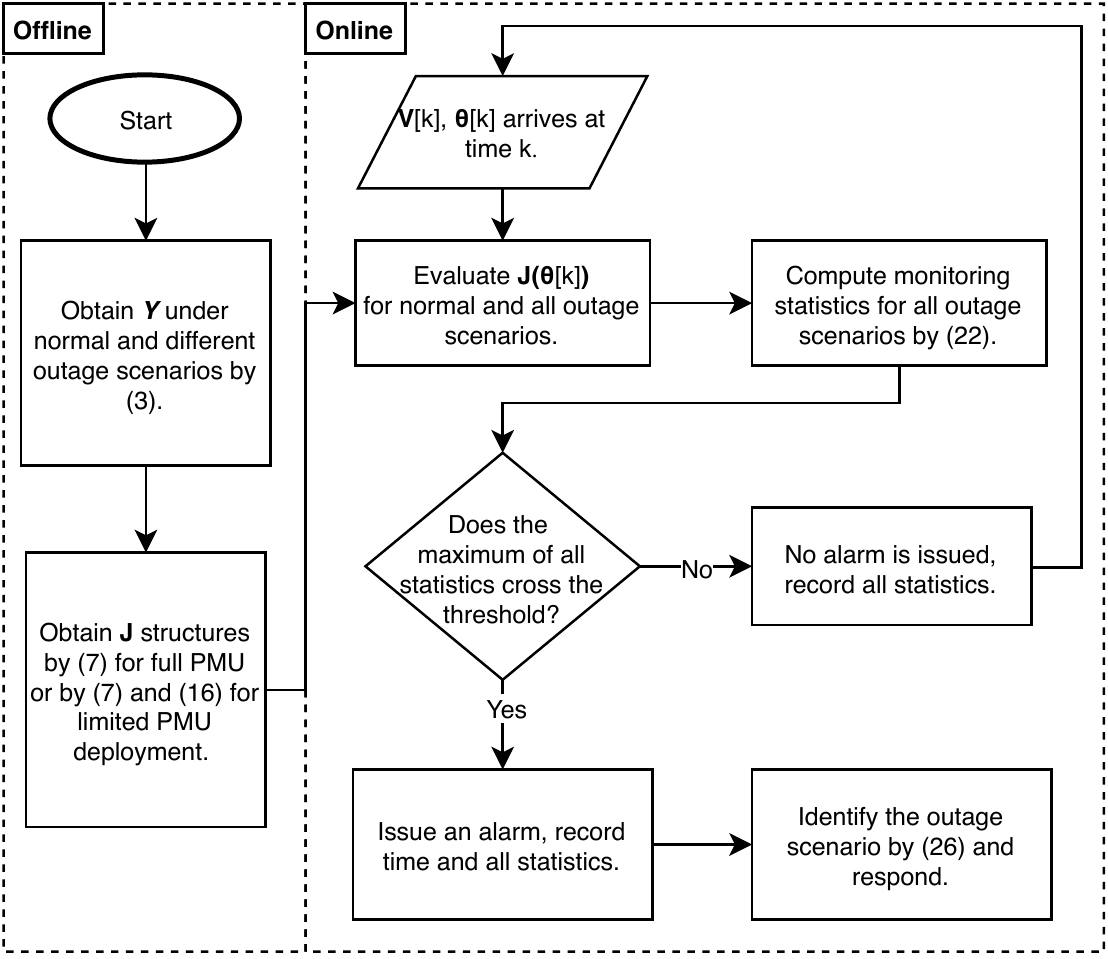}
\caption{\textit{Flowchart summarizing the proposed dynamic outage detection and identification scheme.}}
\label{fig:scheme_flowchart}
\end{figure}

Intuitively, the threshold is crossed when the evidence against the normal condition, i.e., no outage, has accumulated to a significant level. $c$ is a predefined threshold that controls the balance between the detection delay and the false alarm rate. A smaller $c$ corresponds to a more sensitive scheme that may have a quicker detection but could potentially flag more normal fluctuations as outages. One advantage of using the GLR approach is that such trade-off can be systematically quantified. Following \cite{Chen2016}, given a false alarm rate constraint, $c$ could be approximated by
\begin{equation}
\label{eqn:theshold}
c = \ln(ARL_0 \times p) \,,
\end{equation} where $ARL_0$ is the average run length to a false alarm of the scheme when no outage occurs. $p$ is the number of PMUs installed. For example, $c = 18.43$ when $ARL_0 = 1 \text{ day}$ with 39 PMUs installed. With this detection delay and false alarm rate trade-off in mind, ISOs can choose a desired level of sensitivity, catering to the individual system needs, and implement it in the detection scheme through parameter $c$ and $ARL_0$. A flowchart summarizing the working of the detection and identification scheme outlined in this section is shown in Fig. \ref{fig:scheme_flowchart}. 

\renewcommand{\IEEEQED}{\IEEEQEDoff}
\begin{IEEEproof}[Remark 3 (Identification of Tripped Lines)]
Following detection, the actual lines tripped need to be identified so that follow-up, potentially automatic, actions can be taken. 
Since we monitor and compare the likelihood of every outage scenario online, one way to locate the tripped line(s) without any extra computation is to identify the scenarios with the top three likelihoods at the time of detection. In particular, following a detection at time $D$, top-three possible tripped lines can be identified as $\ell_{(1)}, \ell_{(2)}, \text{and } \ell_{(3)}$ such that:
\begin{equation}
\label{eqn:identification}
{W}_{\ell_{(1)}, D} \ge {W}_{\ell_{(2)}, D} \ge {W}_{\ell_{(3)}, D} \ge {W}_{\ell, D}\,,
\end{equation} for all $\ell \in \mathcal{L}$.
\end{IEEEproof}

\section{Case Studies}
\label{sec:results}
\subsection{Simulation Setting}
We test our detection scheme on two IEEE standard test power systems, namely 39 bus New England system \cite{athay1979practical} and 2383 bus Polish system.  System transient responses following an outage are simulated using the open-source dynamic simulation platform COSMIC \cite{Song2016} in which a third-order machine model is used. We conduct extensive single-line outage detection and identification analysis on the 39 bus system by comparing our method to two other methods. Outages on the 2383 bus system are simulated to show that the proposed scheme can be deployed on large-scale systems as well.

We assume that the sampling frequency of PMU is 30 Hz. For every new simulation, we vary the system loads by a random percentage between -5\% and 5\% from the base-line values. Each simulation runs for 10 seconds, and the line outage takes place at the 3rd second. Active power fluctuations are assumed to be uncorrelated and have homogeneous variances where $\sigma^2 = 0.005$ in (\ref{eqn:pre_post_distribution}). Artificial noise is added to all sampled bus angle data, $\Delta \boldsymbol{\theta}$, to account for system and measurement noise \cite{Brown2016}. The noises are drawn from a normal distribution with mean $0$ and standard deviation equivalent to 10\% of the average value of sampled $\Delta{\theta}$ on respective buses. Detection thresholds c in (\ref{eqn:stopping_rule}) corresponding to seven different false alarm rates are obtained by (\ref{eqn:theshold}) and listed in Table \ref{tab:thresholds}. 
\begin{table}
\caption{Detection Thresholds Corresponding to Different Systems and False Alarm Rates}
\label{tab:thresholds}
\centering
\begin{tabular}{cccc}
\hline
\hline
Mean Time to  && Number of PMUs Installed\\
False Alarm (day) & 10 & 39 & 1000 \\
\hline
1/24 & 13.89 & 15.25 & 18.50 \\
1/4 & 15.68 & 17.05 & 20.29 \\
1/2 & 16.38 & 17.74 & 20.98 \\
1 & 17.07 & 18.43 & 21.68 \\
2 & 17.76 & 19.12 & 22.37 \\
7 & 19.02 & 20.38 & 23.62 \\
30 & 20.47 & 21.83 & 25.08 \\
\hline
\end{tabular} 
\end{table}

\subsection{Simulation Results} 
\subsubsection{39 Bus New England System}
The 39 bus system has 39 buses, 10 generators, and 46 transmission lines. We conduct extensive simulation studies for the full PMU deployment and limited PMU deployment scenario. For the latter case, we assume that PMUs are installed on bus 2, 3, 7, 9, 11, 13, 16, 17, 19, and 21. In total, 3000 random simulations of outages at line 1 to 36 are studied, except for line 22 as its outage leads to two separate networks and line 37 to line 46 since they are the only line connecting the generator bus to the system. The proposed method can detect outages instantaneously in most cases with a full PMU deployment. Due to the page constraint, we only present the detection results of a limited PMU deployment here. 

\begin{table}[!t]
\centering
\caption{TIME-STEP BREAKDOWN OF THE DETECTION SCHEME FOR PROCESSING EACH NEW MEASUREMENT}
\label{tab:time_step}
\begin{center}
\begin{tabular}{lll}
\hline
\hline
Step    & Action  & Time Required \\ \hline
0   &   Receive new sample  &   0 \\ \hline
1   &   Evaluate $\mathbf{J}_{0}$ and $\mathbf{J}_{\ell}$ for $\ell \in \mathcal{L}$   &   1 ms \\ \hline
2   &   Compute outage statistics $\mathbf{W}_{\ell}$ for $\ell \in \mathcal{L}$  &   0.227 ms \\ \hline
3   &   Check if $\max \mathbf{W}_{\ell}$ for $\ell \in \mathcal{L}$ exceed $c$  &   0 \\ \hline
\end{tabular}
\end{center}
\end{table}%

\begin{figure}[!t]
\centering
\includegraphics[width=1\linewidth]{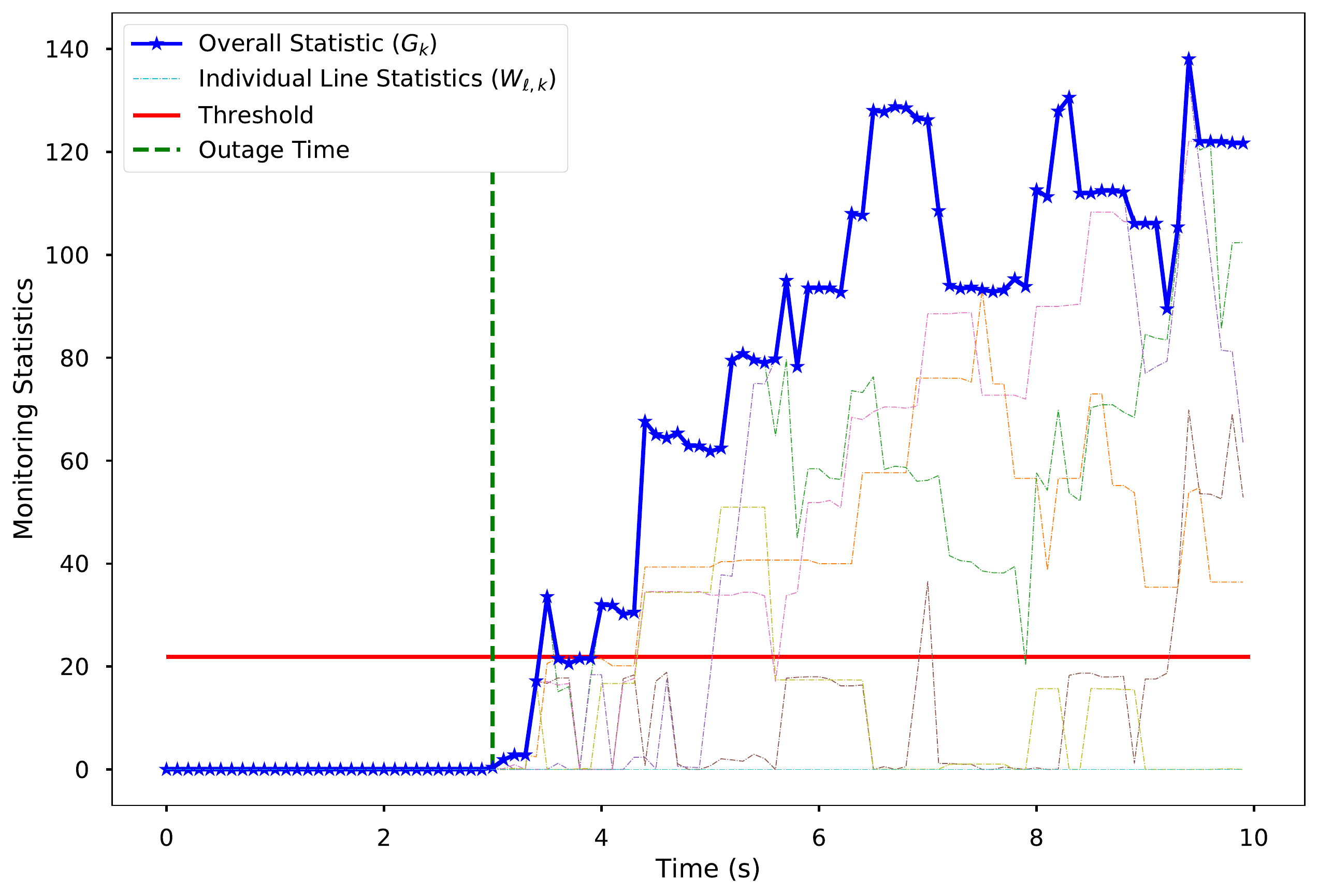}
\caption{\textit{Progression of monitoring statistics for line 10 outage. Individual line statistics are represented by faded dash lines of various colors. Blue solid line is the overall statistic.}}
\label{fig:individual_line10_lim_39}
\end{figure}
We use an outage at line 10 to demonstrate the typical working of the detection scheme. Table \ref{tab:time_step} shows a time-step breakdown for the scheme when processing each new measurement. The execution time is obtained by running the algorithm on a personal laptop with a 2.9 GHz Intel Core i5 processor. Note that a new measurement is collected every 33 ms. Fig. \ref{fig:individual_line10_lim_39} shows the progression of the individual scenario statistics as well as the overall statistic. After the outage (3rd second), individual statistics start to deviate from zero. The overall monitoring statistic rises quickly, too, since it is the maximum of all individual statistics. The scheme issues an alarm when the overall statistic crosses the threshold at time 3.5 seconds. In this case, the scheme records a detection delay of 0.5 seconds. Among all 35 individual statistics representing different outage scenarios, only some have values significantly larger than 0, while most of them stay close to 0 as they are deemed as unlikely scenarios by the detection scheme. 

Also, as we do not have restrictions on the transient stability of the post-outage system, our algorithm does not require bounded signals for outage detection, and it works equally well in stable and unstable scenarios. In fact, an outage that creates an unstable system is easier to detect since it produces stronger signals than those that do not. This is illustrated by a separate simulation example included in the Appendix.

\paragraph{Detection Performance}
Fig. \ref{fig:delay_dist_39_dyn_lim} shows the empirical distribution of detection delays under seven false alarm rates. A more stringent false alarm rate corresponds to a detection scheme with longer delays on average. For example, the scheme with an $ARL_0 = 1/24$ day detects much more outages within 0.25 seconds than the one with $ARL_0 = 30$ days. These differences are not significant. Hence, the proposed scheme's performance based on detection delay is not overly sensitive to different false alarm rates. 
\begin{figure}[!t]
\centering
\includegraphics[width=1\linewidth]{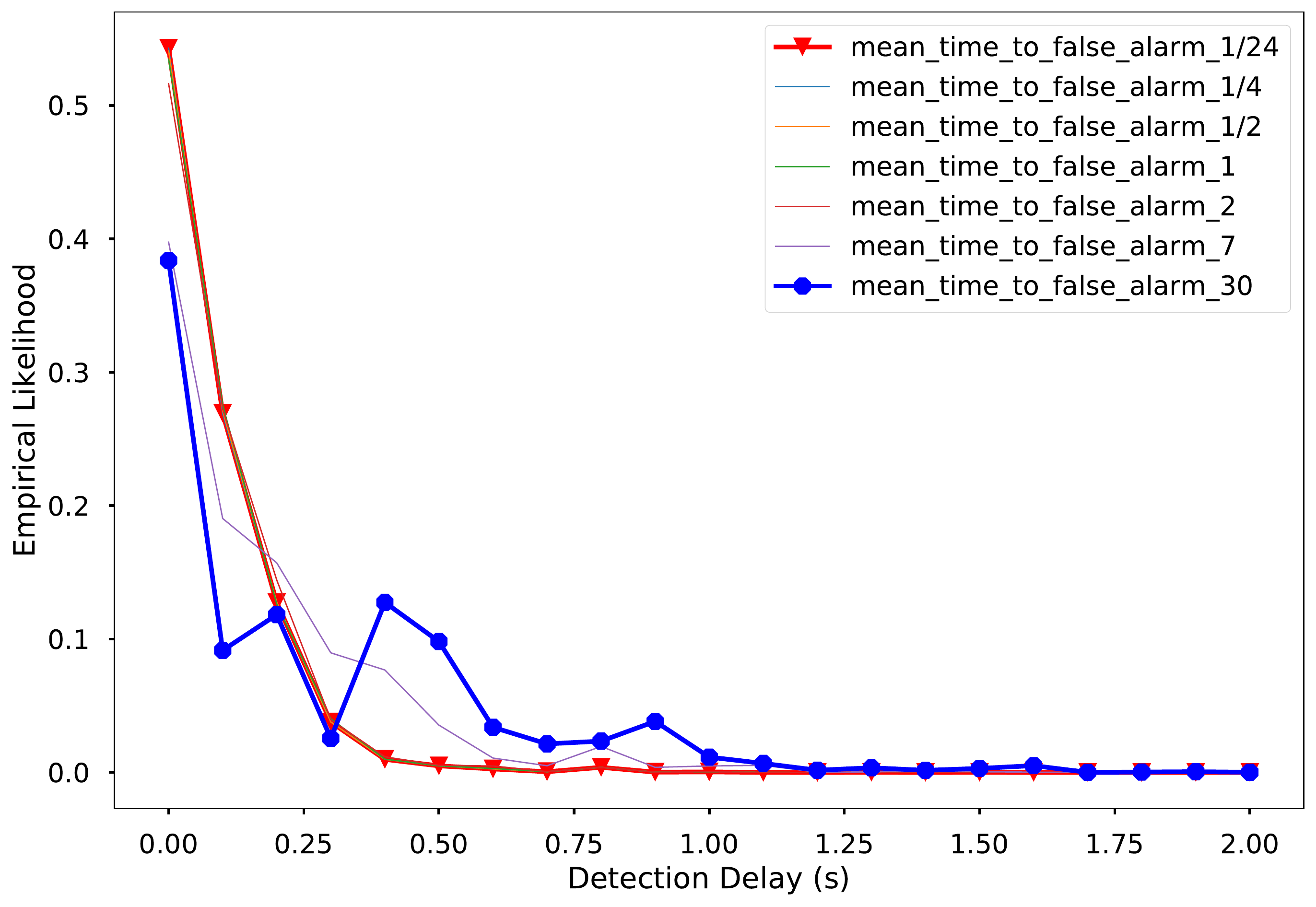}
\caption{\textit{Comparison of the empirical distribution of detection delays in seconds under different false alarm rates. The number in the label is the number of days until a false alarm.}}
\label{fig:delay_dist_39_dyn_lim}
\end{figure}

We have also studied the detection performance across different line outages. There are clear variations in terms of detection delay among those detected outages. These variations can be largely attributed to the PMU placement and the grid topology.
\begin{figure}[!t]
    \centering
  \subfloat[\label{fig:boxplot_delay_30day_pmu}]{%
       \includegraphics[width=1\linewidth]{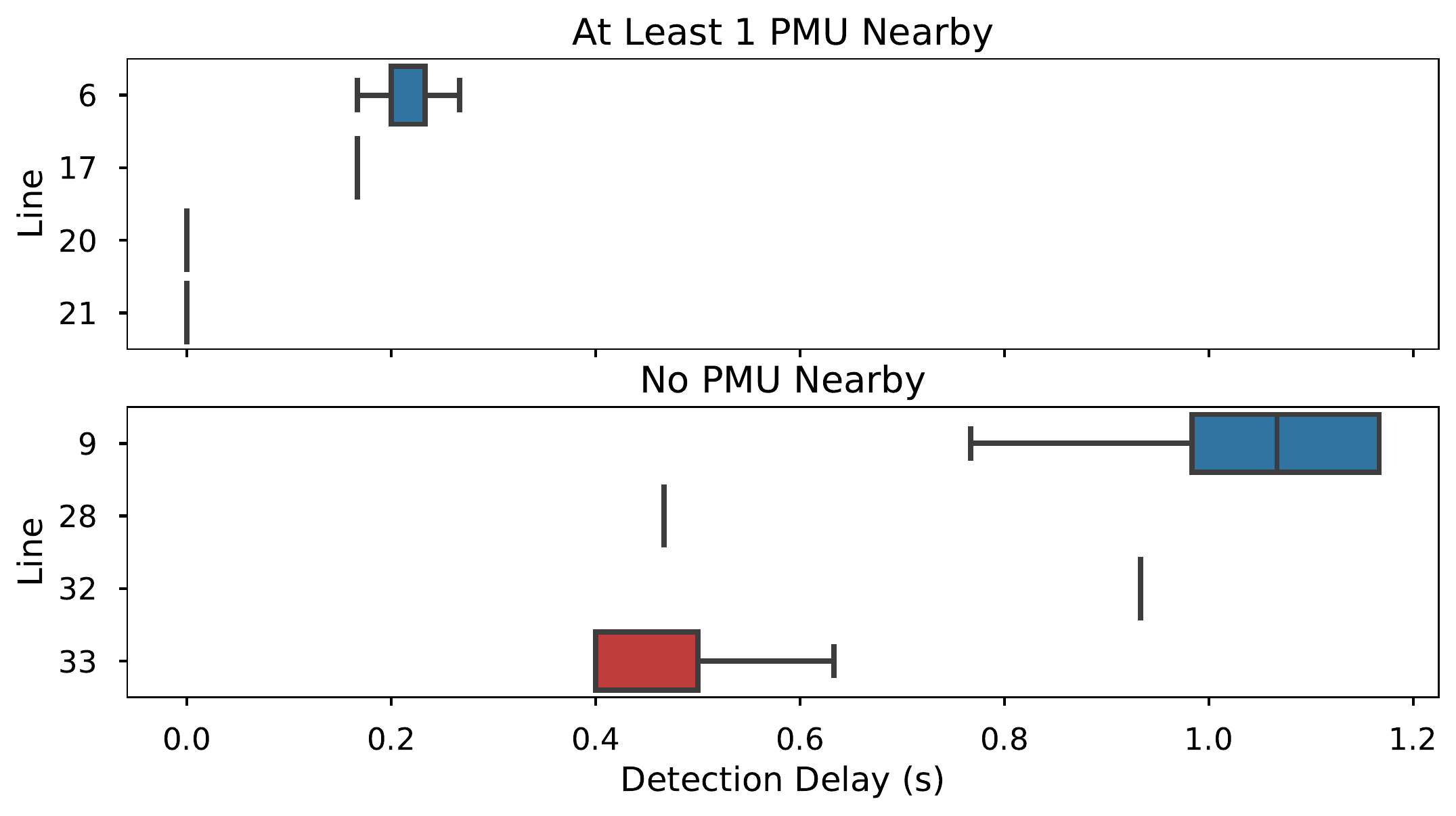}}
    \\
  \subfloat[\label{fig:boxplot_delay_30day_critical}]{%
        \includegraphics[width=1\linewidth]{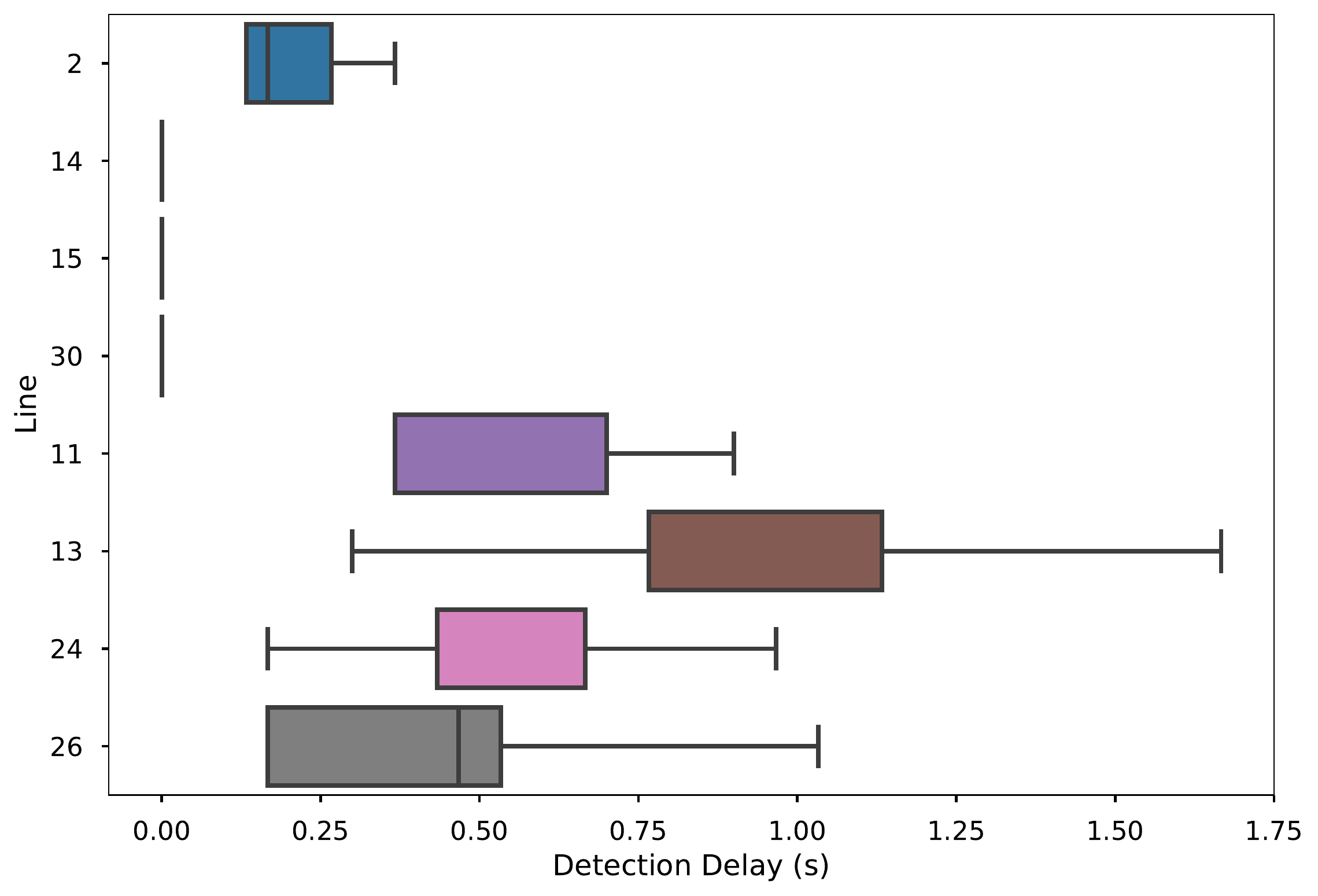}}
  \caption{\textit{Boxplot of the empirical distributions of detection delay in seconds for (a) lines with at least 1 PMU nearby and those without,  (b) lines at different topological locations.}}
  \label{boxplot_delay_30day_critical} 
\end{figure}
For outages with almost zero detection delay, they are lines where either PMUs are installed on both ends of the line, e.g., line 3, 21, and 23, or one PMU is connected to the line, e.g., line 20, 25, and 27. Signals can be readily picked up by nearby PMUs. On the other hand, the absence of PMU nearby may have contributed to the longer detection delays. In particular, there are no PMUs available on either end of line 9, 10, 28, 32, 33, and 34. These outage signals have to be detected by sensors far away from the location. Fig. \ref{fig:boxplot_delay_30day_pmu} summarizes the comparison.

Another factor is the power grid topology. The scheme recorded shorter delays for line 2, 14, 15, and 30. It is observed that these outages produced severe disturbances. Line 2, 14, and 15 connect to a generator bus, and line 30 connects a subnetwork to the main network. On the other hand, outages at line 5, 11, 13, and 26 produced weaker and shorter disturbances, which are more difficult to detect. Consequently, they recorded longer detection delays. See Fig. \ref{fig:boxplot_delay_30day_critical} for the comparison.

\paragraph{Comparison with Other Methods}
\begin{table}
\caption{Comparison of Detection Delay (s) of Three Different Line Outages Under Different Detection Schemes}
\label{tab:delay_comparison_39}
\centering
\begin{tabular}{llcccc}
\hline
\hline
	& & \multicolumn{4}{c}{Mean Time to False Alarm (day)} \\ 
Line	& Scheme & 1/24 & 2 & 7 & 30 \\ \hline
 26		&	DC - full & 9.9908 & 9.9908 & 9.9908 & 9.9908 \\ 
		&	DC - limited & \textendash & \textendash & \textendash & \textendash \\ 
		&	Ohm's Law - limited & 2.8150 & 3.0963 & 3.1406 & 3.9333 \\ 		
		&	AC - limited & 0.1001 & 0.1005 & 0.3300 & 0.3489 \\ \hline
 27		&	DC - full & 4.5398 & 4.5398 & 4.5398 & 4.5398 \\ 
		&	DC - limited & \textendash & \textendash & \textendash & \textendash \\ 
		&	Ohm's Law - limited & 3.3044 & 3.5000 & 3.6900 & 3.8630 \\ 	
		&	AC - limited & 0.0012 & 0.0012 & 0.0026 & 0.0039 \\ \hline
 34		&	DC - full & 0.1801 & 0.1801 & 0.1801 & 0.1801 \\ 
		&	DC - limited & \textendash & \textendash & \textendash & \textendash \\ 
		&	Ohm's Law - limited & 1.5811 & 2.9250 & 3.2014 & 3.6788 \\ 	
		&	AC - limited & 0.0879 & 0.0879 & 0.1558 & 0.4994 \\ \hline
\end{tabular}
\end{table}
We compared the proposed method's outage detection performance with two other methods. The line outages considered here are line 26, 27, and 34. Other methods considered here are the static detection method based on the DC power flow model in \cite{Chen2016}, under a full and limited PMU deployment, and the CUSUM-type central rule based on Ohm's law in \cite{Jamei2017a}, with a limited PMU deployment. The placement of 10 PMUs is the same for all methods. For the CUSUM scheme in \cite{Jamei2017a}, parameters are chosen to satisfy the same false alarm rates in Table \ref{tab:thresholds} based on formula in \cite{montgomery2007introduction}. The respective detection delays are summarized in Table \ref{tab:delay_comparison_39}. A dash means a missed detection. It can be seen that our proposed method, ``AC - limited'', is consistently faster at detecting outages than the other methods.

\begin{figure}[!t]
    \centering
  \subfloat[39 PMUs\label{fig:identification_full_top3}]{%
       \includegraphics[width=.5\linewidth]{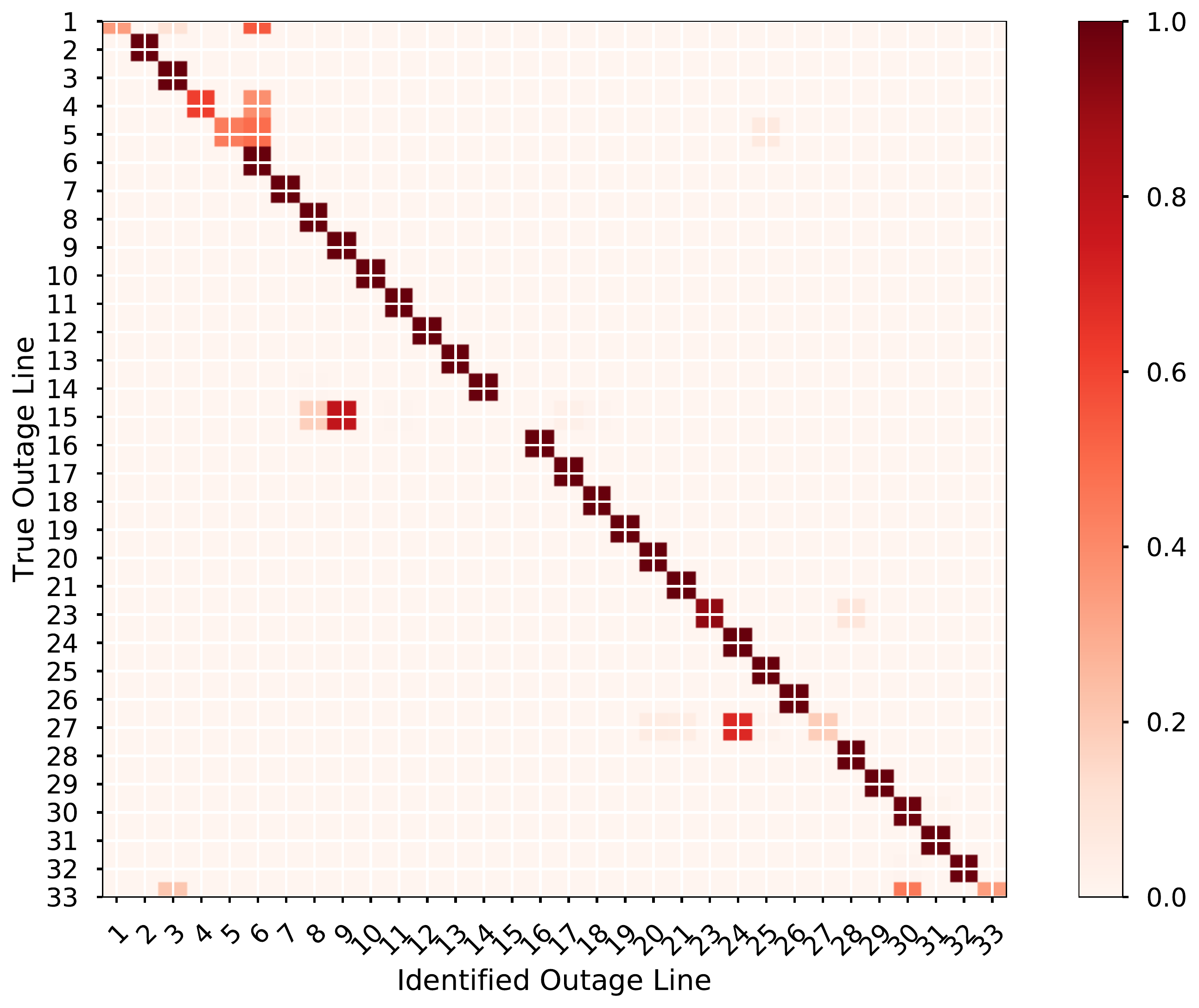}}
  \subfloat[10 PMUs\label{fig:identification_top3_tree_10pmu}]{%
       \includegraphics[width=.5\linewidth]{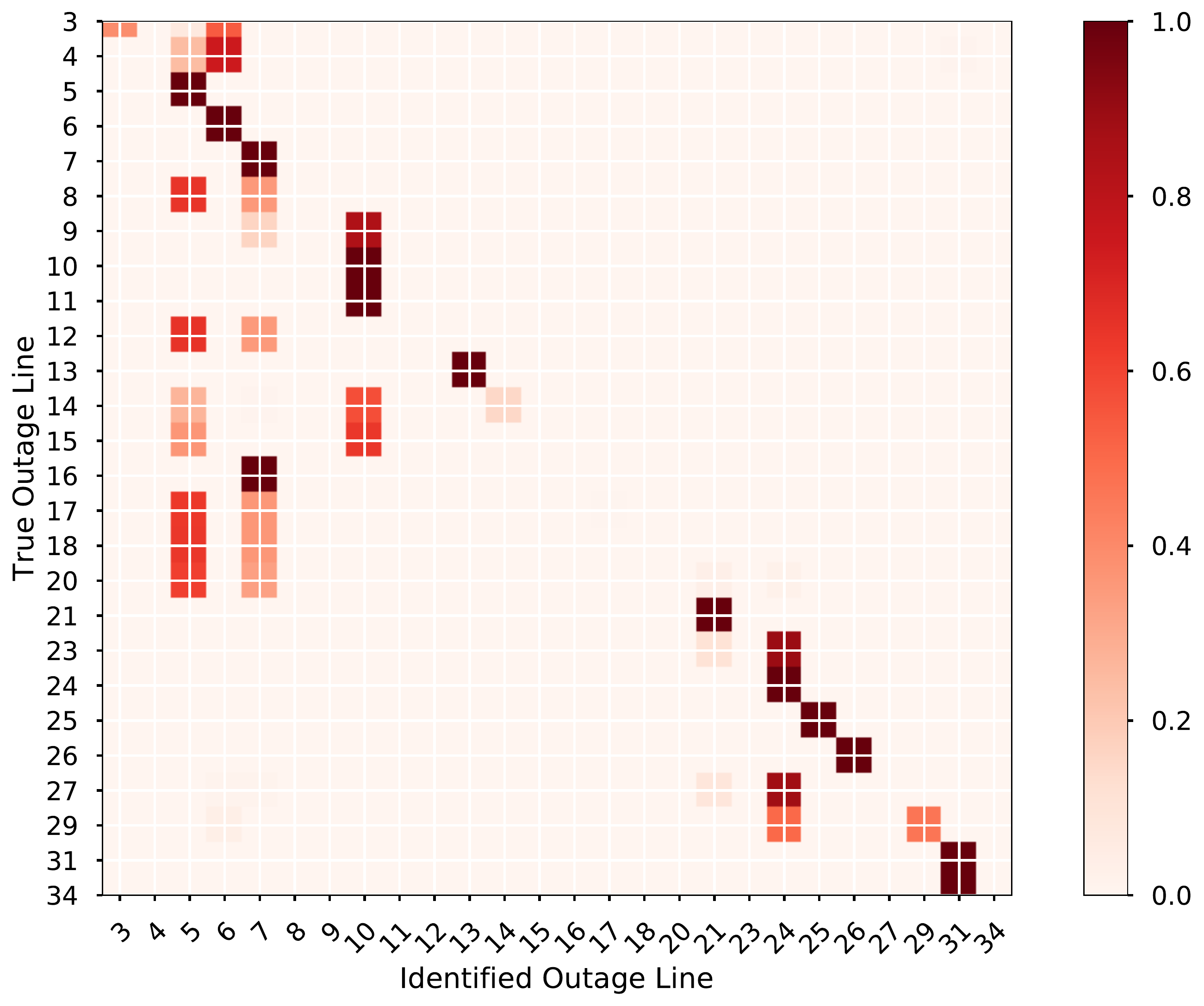}}
  \caption{\textit{Heat map showing the identification accuracy of the proposed method in the 39 bus system with (a) a full PMU deployment and (b) 10 PMUs deployed.}}
  \label{fig:identification_heatmap} 
\end{figure}

\paragraph{Identification Performance}
We analyzed the identification performance by comparing the true outage line with the identified line. The results are shown in Fig. \ref{fig:identification_heatmap}. True outage lines are listed on the vertical axis, and the lines identified are on the horizontal axis. Cell color represents the empirical likelihood of identification of different lines. Therefore, a perfect identification scheme would have all diagonal cells equal to 1 and 0 everywhere else. As seen from the figure, most lines can be accurately identified. When the scheme misses the true outage line, it often misidentifies the adjacent line as tripped. This suggests that a way to improve the effectiveness of corrective actions following an outage is to inspect the identified line as well as its neighboring lines. 

\subsubsection{2383 Bus Polish System}
To show that the proposed dynamic detection scheme can be deployed in a system with realistic network size, outages in the 2383 bus system are studied. This test system has 2383 buses and 2896 transmission lines. 1000 PMUs are assumed to be placed at randomly selected locations in the system. Eight different line outages are simulated to test the proposed detection scheme. Detection delay results corresponding to four different false alarm rates are reported in Table \ref{tab:delay_2383}. Considering the size of the system, detecting a single-line outage is much more difficult. Therefore, delays experienced are considerably longer than those in the 39 bus system. There are also several undetected outages. 
\begin{table}
\caption{Detection Delay (s) of Eight Different Line Outages in 2383 Bus System with 1000 PMUs Deployed}
\label{tab:delay_2383}
\centering
\begin{tabular}{lcccc}
\hline
\hline
 	& \multicolumn{4}{c}{Mean Time to False Alarm (day)} \\ 
Line	& 1/24 & 2 & 7 & 30 \\ \hline
600	& 4.6667 & 4.6667 & 4.6667 & 4.6667\\ 
700	& 1.3667 & 1.3667 & 1.3667 & 1.3667 \\ 
750	& 4.9000 & 4.9000 & 4.9000 & 4.9000 \\ 
800	& 1.3667 & 1.3667 & 6.7667 & 6.7667 \\ 
900	& \textendash & \textendash & \textendash & \textendash \\ 
1000	& \textendash & \textendash & \textendash & \textendash \\ 
1050	& 1.3667 & 1.3667 & 1.3667 & 1.3667 \\ 
1650	& \textendash & \textendash & \textendash & \textendash \\ \hline
\end{tabular}
\end{table}

\section{Conclusion}
\label{sec:conclusion}
In this work, we developed a real-time dynamic line outage detection and identification scheme based on the AC power flow model and GLR scheme. We derived a time-variant small-angle relationship between bus voltage angles and active power injections. We obtained the pre- and post-outage statistical models of the angle variations. The proposed scheme is effective in both detection and identification. It is also scalable, as seen from the results in the 2383 bus system. 

For further research, we would investigate the optimal number and placement of a limited number of PMUs. As seen from Section \ref{sec:results}, there is a varying level of detection delays due to PMU placement. The number of PMUs needed to achieve a certain level of identification accuracy is also worth investigating. We would also consider incorporating generator dynamics into our system model, where we hope the detailed physical model could provide an even better direction for outage detection and identification.

\begin{figure}[t]
\centering
\includegraphics[width=1\linewidth]{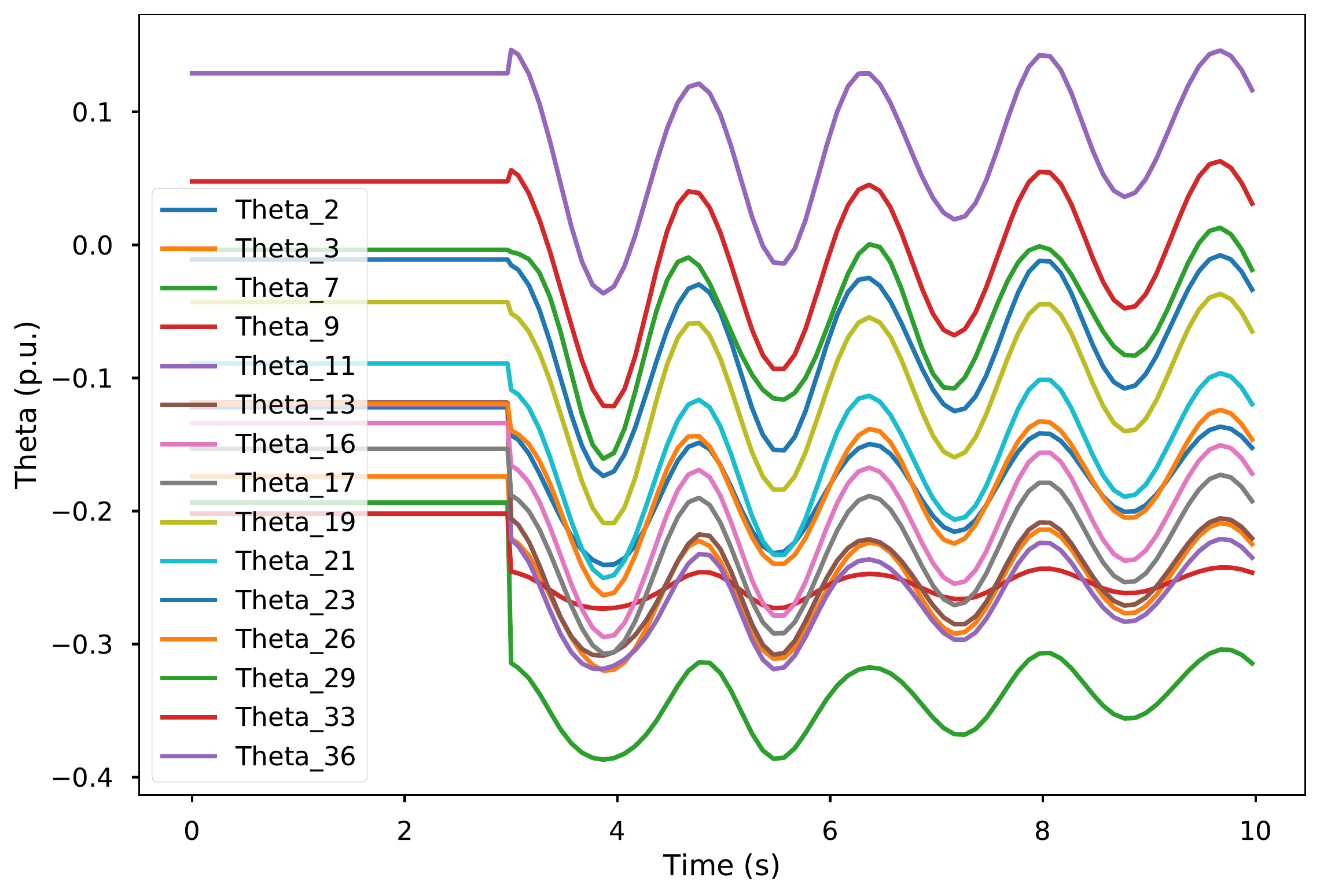}
\caption{\textit{The progression of bus voltage phase angles after the outage of line 37. Each line represents the voltage phase angles from one of the buses.}}
\label{fig:line_37_outage}
\end{figure}
\begin{figure}[t]
\centering
\includegraphics[width=1\linewidth]{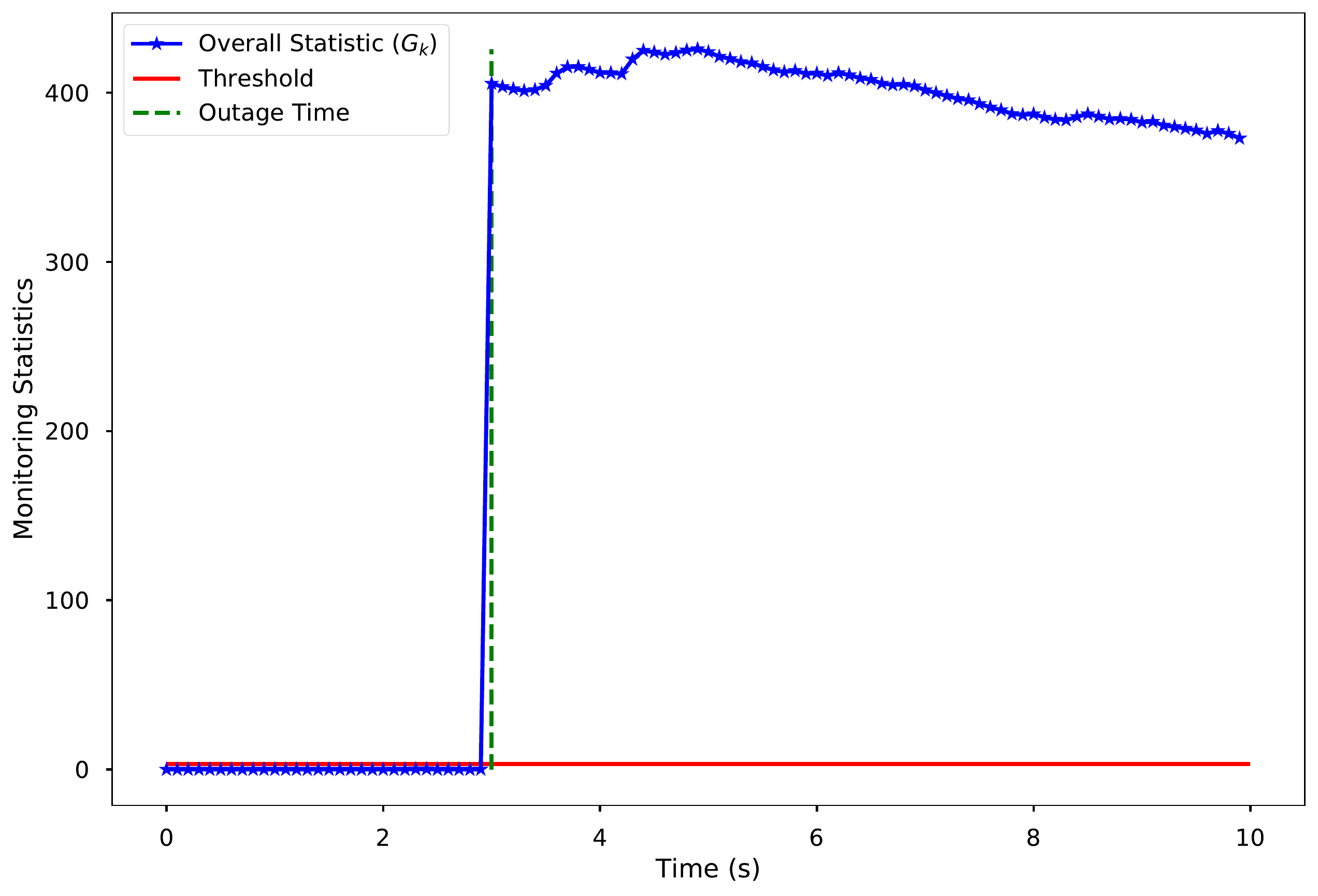}
\caption{\textit{The progression of the monitoring statistic for line 37 outage.}}
\label{fig:line_37_detection}
\end{figure}

\appendix[Unstable Post-Outage System]
\label{sec:appendix}
We show a simulation example to illustrate the working of the detection scheme when the outage creates an unstable and transient system. In the 39-bus system, line 37 outage creates large disturbances throughout the system, as shown in Figure \ref{fig:line_37_outage}. From the onset of the outage to the end of the simulation, voltage phase angles at most buses show no significant sign of stabilization. The detection scheme is able to detect the outage immediately, as shown in Figure \ref{fig:line_37_detection}. In this case, the monitoring statistic records a significantly large value, indicating that the strength of the signals is strong. 


\ifCLASSOPTIONcaptionsoff
  \newpage
\fi



%
\begin{IEEEbiography}{Xiaozhou Yang}
received the B.S degree in industrial and systems engineering from National University of Singapore, Singapore, in 2017. 
He is currently a Ph.D. student in the Department of Industrial Systems Engineering and Management of the same university. He is also a researcher at the Future Resilient Systems programme of Singapore-ETH Centre. 
His research interests include advance data analytics in power system condition monitoring, real-time outage detection, and identification.
\end{IEEEbiography}

\begin{IEEEbiography}{Nan Chen}
received the B.S. degree in automation from Tsinghua University, Beijing, China, in 2006, the M.S. degree in computer science in 2009, and the M.S. degree in statistics and the Ph.D. degree in industrial engineering from the University of Wisconsin-Madison, Madison, WI, USA, both in 2010.
He is currently an Associate Professor with the Department of Industrial Systems Engineering and Management, National University of Singapore, Singapore. His research interests include statistical modeling and surveillance of engineering systems, simulation modeling design, condition monitoring, and degradation modeling.
\end{IEEEbiography}

\begin{IEEEbiography}{Chao Zhai}
(S'12-M'14) received the Bachelor's degree in automation engineering from Henan University in 2007 and earned the Master's degree in control theory and control engineering from Huazhong University of Science and Technology in 2009. He received the PhD in complex system and control from the Institute of Systems Science, Chinese Academy of Sciences, Beijing, China, in June 2013. From July 2013 to August 2015, he was a Post-Doctoral Fellow with the University of Bristol, Bristol, UK. Since February 2016, He has been a Research Fellow with Singapore-ETH Centre and Nanyang Technological University, Singapore. His current research interests include cooperative control of multi-agent systems, power system stability and control, social motor coordination, and evolutionary game theory.
\end{IEEEbiography}


\vfill


\end{document}